\documentclass[twocolumn,aps,amssymb,prl,showpacs]{revtex4}
\usepackage{latexsym}
\usepackage{graphicx}
\usepackage{color}

\begin{document}

\title{ Slow conductance relaxation in insulating granular Al: evidence for screening effects }
\author{J. Delahaye, J. Honor\'e, T. Grenet}
\address{Institut N\'eel, CNRS $\&$ Universit\'e Joseph Fourier, BP 166, 38042 Grenoble, France}

\date{\today} 

\begin{abstract}

It is shown that the conductance relaxations observed in electrical field effect measurements on granular $Al$ films are the sum of two contributions. One is sensitive to gate voltage changes and gives the already reported anomalous electrical field effect. The other one is independent of the gate voltage history and starts when the films are cooled down to low temperature. Their relative amplitude is strongly thickness dependent which demonstrates the existence of a finite screening length in our insulating films and allows its quantitative estimate (about $10nm$ at $4K$). This metallic-like screening should be taken into account in the electron glass models of disordered insulators.

\end{abstract}
 \bigskip
\maketitle


During the last 20 years, slow and glassy conductance relaxations were found in several disordered insulating systems \cite{FirstExpElectronGlass,OvadyahuPRB02,GrenetEPJB07}. These relaxations have been revealed and often studied by means of field effect measurements. In such experiments, disordered insulating films are used as (weakly) conducting channels of MOSFET devices which allow us to measure their conductance response to gate voltage ($V_g$) changes. After a quench at e.g. $4K$, a slow and endless decrease of the conductance is found as long as $V_g$ is kept constant, and any $V_g$ change triggers a new conductance relaxation. Moreover, the system keeps some memory of its $V_g$ history: Any stay under a fixed $V_g$ value remains printed for some time in $G(V_g)$ sweeps as a conductance dip centered on this value. We have also shown recently in granular $Al$ thin films that these $V_g$ induced relaxations display aging, i.e. the dynamics of the system depends on the time spent at low temperature, a characteristic property of glassy systems \cite{GrenetAging10}.

 Several experimental findings suggest that these slow conductance relaxations could reflect the properties of the "electron glass" \cite{OvadyahuPRL98,OvadyahuElectronGlass}, a glassy state theoretically predicted in the 1980s for disordered insulators \cite{DaviesPRL82,ElectronGlass80}. According to theoretical and numerical studies, a system of electrons with ill-screened interactions and disorder will need an infinite time to reach its equilibrium state at low temperature, the relaxation towards equilibrium being characterized by a conductance decrease of the system \cite{RecentEGTheory}. Up to now, the electron glass problem has received considerable theoretical developments but only few experimental illustrations, which explain the interest in the electrical field effect results. Moreover, most of the experiments have focused on the electrical conductance and its $V_g$-induced relaxations but less is known concerning the dielectric properties and their possible time evolution at low $T$.

Since even after a long stay under a fixed $V_g$ the conductance decrease shows no sign of saturation, it is not possible to define the conductance relaxation relative to the equilibrium value, which is unknown. Instead, short excursions to $V_g$ values never explored before are often used to define a $V_g$ history-free reference conductance $G_{ref}$ \cite{OvadyahuPRL98,OvadyahuPRB02,GrenetEPJB07,GrenetAging10}. This history-free conductance was sometimes called the "off-equilibrium" conductance since the system has never been allowed to equilibrate at these $V_g$s. In the present Letter, we show that $G_{ref}$ is not constant in time after a cool down to $4.2K$ in granular $Al$ films thicker than $10nm$. We also show that this feature demonstrates the existence of a screening length in our insulating granular films and allows its quantitative estimate.

 Our granular $Al$ films were prepared by e-gun evaporation of $Al$ under a partial pressure of $O_2$, as described elsewhere \cite{GrenetEPJB07}. By changing the $O_2$ pressure, we can tune the resistance of the films from metallic to insulating. For insulating samples, x-rays and TEM studies have revealed an assembly of crystalline $Al$ grains with a typical size of a few nanometers. They are believed to be separated by thin insulating $Al_20_{3-x}$ layers. MOSFET devices are made by deposition of granular $Al$ films on top of heavily doped $Si$ wafers (the gate) covered by a $100nm$ thick thermally grown $SiO_2$ layer (the gate insulator).

The film conductance $G$ was measured by using a two terminal ac technique, employing a FEMTO current amplifier DLPCA 200 and a lock in amplifier SR 7265. Source-drain voltage was such that $G$ stays in the Ohmic regime. All the electronic equipments sensitive to room temperature drifts were placed in a thermalized chamber with a $T$ stability better than $0.1K$. This last point was crucial in order to follow $G$ variations with a precision of $10^{-3}\%$ during weeks of measurements.

For $4.2K$ measurements, the MOSFET devices are mounted in a box filled with He exchange gas and plunged into a $100l$ liquid He dewar. The sample temperature follows the $mK$ variations of the liquid He bath around $4.2K$ and a carbon glass thermometer close by was used to correct these temperature variations. The time required to cool down the sample from room temperature to $4.2K$ is about $10min$.

\begin{figure}
    \includegraphics[width=8cm]{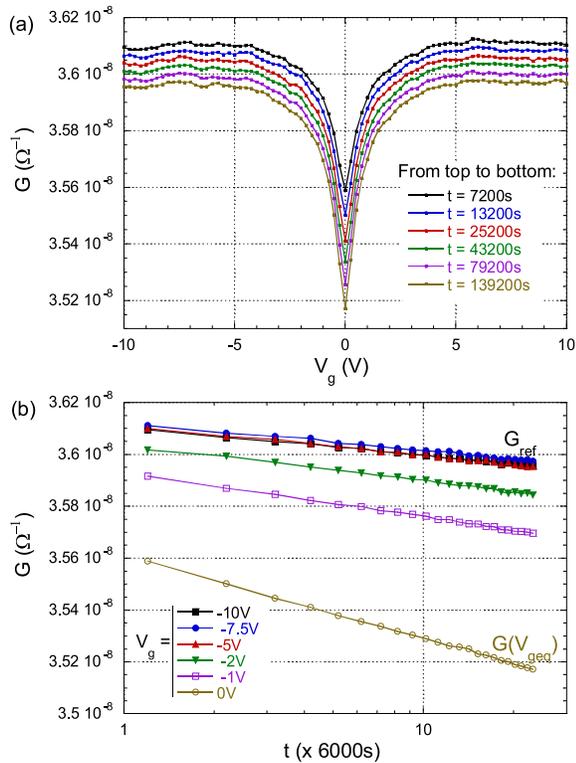}
    \caption{(a) $G(V_g)$ curves measured at different times $t$ after a cool down to $4.2K$. $V_g=V_{geq}=0V$ between $V_g$ sweeps. (b) Corresponding $G(t)$ curves for negative $V_g$s in and out of the conductance dip region. The sample was $20nm$ thick with $R_{\Box} = 550M\Omega$ at $4.2K$. See the text for the details.}\label{Figure1}
    \end{figure}

 For Fig. \ref{Figure1}a, a MOSFET device with a $20nm$ thick granular $Al$ channel was cooled down to $4.2K$ and maintained at this temperature under $V_{geq} = 0V$. Fast $V_g$ sweeps from $-15$ to $+15V$ ($250s$ long) were taken every $6000s$ after the cool down. All the $G(V_g)$ curves display a conductance dip centered on $V_{geq}$ which reflects the memory of the relaxation associated with the stay under $V_{geq}$. Far enough from $V_{geq}$ (here for $|V_g| > 5V$), $G$ is roughly constant and this baseline reflects the "off-equilibrium" conductance $G_{ref}$ previously mentioned. Looking at the time evolution of $G(V_g)$ curves, two features are salient. First, the amplitude of the dip increases as a function of time. This result is well known from previous studies \cite{OvadyahuPRL98,OvadyahuPRB02,GrenetEPJB07,GrenetAging10}: the longer the stay under $V_{geq}$, the more pronounced the dip is. Second, and this is the new feature we aim to discuss here, the baseline conductance also decreases as a function of time. As highlighted in Fig. \ref{Figure1}b, the $G$ decrease is well described at any $V_g$s by a $\ln t$ dependence, more pronounced in the dip region because of the superposed baseline relaxation and dip growth.

To be more quantitative, we can define a parameter whose physical meaning will become clear later, the slope ratio $SR$ of the $G$ relaxation slope in Fig. \ref{Figure1}b at $V_{geq}$ to the $G_{ref}$ relaxation slope (at $-10V$ for example). For the sample of Fig. \ref{Figure1}, $SR \simeq 2.9$. We have measured 4 different $20nm$ thick films with $R_{\Box}$ values from $10M\Omega$ to $10G\Omega$. A baseline relaxation was always observed, with $SR$ at $4K$ between $2$ and $3$. For one film, the gate and the gate insulator were respectively $Al$ and alumina and consistent results were obtained which excludes any role of a specific gate insulator material. Detailed investigation of the vertical electrical homogeneity of the films has also shown that the baseline relaxation is not related to a specific layer close to the film-substrate interface \cite{SupplementaryMaterial}.

We checked that this baseline relaxation was independent of the $V_g$ history by using the "two dip" protocol \cite{OvadyahuPRL98}. After some time under $V_{geq1}$, $V_{geq}$ was changed to $V_{geq2}$ for the rest of the experiment. We observed the formation of a new dip at $V_{geq2}$ and the erasure of the old one at $V_{geq1}$ (as already known \cite{OvadyahuPRL98,GrenetEPJB07}), but the baseline relaxation in any $V_g$ range far enough from $V_{geq1}$ and $V_{geq2}$ continue as if no $V_{geq}$ change was imposed.

One may think that the baseline relaxation is induced by the $V_g$ sweeps themselves. Indeed, when measuring $G(V_g)$ curves, a small dip starts to form at each measured point. The baseline thus reflects a short time relaxed $G$ value ($10s$ being typical for our sweep parameters). We know from the "two dip" protocol results that a dip formed during $10s$ will be erased roughly in about $100s$ \cite{OvadyahuPRB02,GrenetEPJB07}. In our case, we wait for $6000s$ between two $V_g$ sweeps, thus no memory of the previous sweep is expected out of the dip region. In one experiment, we checked that the baseline relaxation was unaffected by the suspension of $V_g$ sweeps during $30h$.

An experimental artifact that may explain the baseline relaxation is the existence of a $T$ drift after the cool down. Because of the rapid divergence of the films resistance at low $T$, the observed baseline variations correspond typically to drifts of few $mK$ ($\simeq 2.5mK$ for the whole relaxation observed at $4.2K$ in Fig. \ref{Figure1}). We have performed different tests (thermometer stability, comparison with and without $He$ exchange gas) which indicate that the $T$ drifts are much smaller than the baseline relaxation amplitude. The effect of the film thickness discussed below will definitely ruled out any interpretation in terms of a $T$ drift.

If the baseline relaxation is a property of the granular film itself, how can we explain it?
A simple hypothesis is to state that the granular $Al$ film, although electrically insulating, has a metallic-like screening length $L_{sc}$ that is smaller than its thickness $T_h$ ($T_h=20nm$ in Fig. \ref{Figure1}). Then, only the layer of the film located at a distance smaller than $L_{sc}$ from the gate insulator is sensitive to $V_g$ changes. The conductance dip $\Delta G_{dip}$ reflects the relaxation of this layer, whereas the conductance baseline $G_{ref}$ reflects the relaxation of the rest of the film. The conductance relaxation measured at $V_{geq}$ is then the sum of the two contributions: $\Delta G(V_{geq},t) = \Delta G_{dip}(t)+\Delta G_{ref}(t)$. Since the relaxations at any $V_g$ are well described by a $\ln t$ dependence, we can write $\Delta G_{ref}(t) = -A_{ref}\ln t$ and $\Delta G_{dip}(t)=-A_{dip}\ln t$. The slope ratio $SR$ is by definition equal to $(A_{dip}+A_{ref})/A_{ref}$. If the film is homogeneous, it is natural to suggest that $A_{dip}$ and $A_{ref}$ are respectively proportional to $L_{sc}$ (the $V_g$ sensitive layer thickness) and $T_h - L_{sc}$ (the $V_g$ insensitive layer thickness). Then, $SR=T_h/(T_h-L_{sc})$ as long as $T_h>L_{sc}$. The results for the $20nm$ films ($SR$ between 2 and 3) give estimates for the screening length $L_{sc}$ between $10$ and $13nm$.

\begin{figure}
    \includegraphics[width=7.5cm]{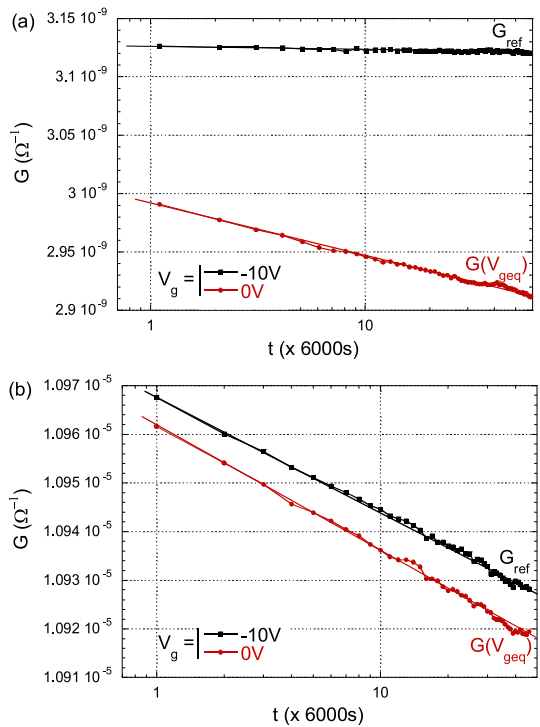}
    \caption{$G(t)$ measurements after a cool down at $4.2K$ for: (a) a $10nm$ thick film, $R_{\Box}=6G\Omega$; (b) a $100nm$ thick film, $R_{\Box} =2M\Omega$. The $G(t)$ relaxations are plotted at $0V$ ($V_{geq}$) and at $-10V$ (baseline relaxation $G_{ref}$).}\label{Figure3}
    \end{figure}

If our simple model is correct, then the baseline relaxation relative to that measured at $V_{geq}$ must change with the thickness of the films. This is illustrated in Fig. \ref{Figure3}, where the experiment of Fig. \ref{Figure1} was reproduced for two granular $Al$ films $10nm$ and $100nm$ thick. A clear thickness dependence is visible. Qualitatively, the baseline relaxation is almost absent for the $10nm$ thick film (Fig. \ref{Figure1}a, $SR=20$) while it is almost equal to that measured at $V_{geq}$ for the $100nm$ thick film (Fig. \ref{Figure1}b, $SR=1.1$). The agreement with the above simple picture is also quantitative. These $SR$ values gives $L_{sc}$ estimates of about $10nm$, close to the results obtained from $20nm$ thick films ($10-13nm$). We stress that the $SR$ variations observed for different thicknesses are much more important than those observed for a given thickness and different $R_{\Box}s$.

\begin{figure}
    \includegraphics[width=9cm]{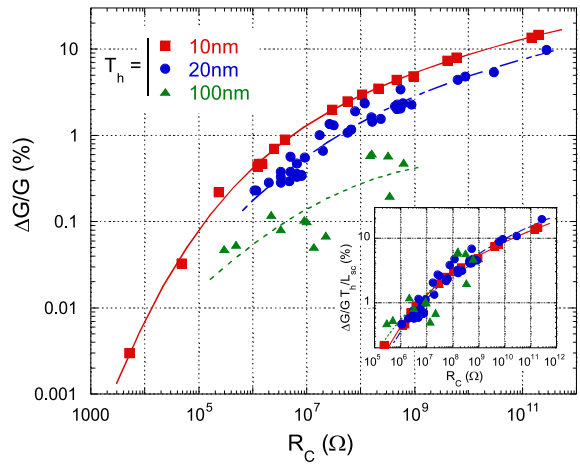}
    \caption{$\Delta G/G$ values measured $20h$ after a cool down at $4.2K$ as a function of $R_C$ (see the text for the details). The films are $10nm$ (squares), $20nm$ (circles) and $100nm$ (triangles) thick. The lines are guides for the eye. In the inset, $(\Delta G/G) \times (T_{h}/L_{sc})$ versus $R_C$. Typical error bars on each point are about 10\%.}\label{Figure4}
    \end{figure}

Our simple model also predicts a thickness dependence for the amplitude of the conductance dip $\Delta G/G=[G_{ref}-G(V_{geq})]/G_{ref}$. $\Delta G/G$ data, measured for $10nm$, $20nm$ and $100nm$ thick films are shown in Fig. \ref{Figure4} as a function of their percolation critical resistance $R_C$ \cite{ESBook84}. In 2D films ($T_h$ less than the percolation radius $L_0$), $R_C\simeq R_{\Box}$, but for thicker ones, $R_C\simeq (T_h/L_0) R_{\Box}$. Taking $L_0 \simeq 20nm$ \cite{DelahayeEPJB08}, we estimate $R_C$ as $R_{\Box}$ for $10$ and $20nm$ films and $5R_{\Box}$ for $100nm$ films. All the samples were measured after being kept under $V_{eq}=0 V$ for $t=20h$ after a cool down at $4.2K$. As can be seen and was already reported \cite{FirstExpElectronGlass,GrenetEPJB07}, $\Delta G/G$ increases significantly with $R_C$. Besides, a clear reduction of $\Delta G/G$ is visible for a given $R_C$ as $T_h$ increases, in qualitative agreement with the fact that the conductance dip originates from a layer of a fixed thickness. The agreement is also quantitative. As long as $T_h < L_{sc}$, $\Delta G/G(t)$ does not depend on $T_h$. Let us note its value $C(t)$ (of course $C(t)$ depends on $R_C$). When $T_h > L_{sc}$, one has $\Delta G /G(t) = A_{dip} \ln t/(G_0-A_{ref} \ln t) \simeq A_{dip} \ln t/G_0= C(t) L_{sc}/T_h$ ($\Delta G/G(t)\ll 1$). Then, plotting $(T_h/L_{sc})\times(\Delta G/G)$ for all the samples should give one single curve $C(t,R_C)$. This is confirmed in the inset in Fig. \ref{Figure4} for which we have assumed a screening length of $10nm$ for all the films.

Finally, our model provides a natural explanation for the increase of the $\Delta G/G$ data scatter with the thickness, as observed in Fig. \ref{Figure4}. Small variations of the $O_2$ pressure or of the $Al$ evaporation rate are always present during the deposition of granular $Al$ films. Thus in thick samples, the measured $R_{\Box}$ (and the estimated $R_C$) does not necessarily reflect the resistance of the unscreened $10nm$ thick layer deposited on top of the gate insulator which determines the amplitude of the dip. The quality of the scaling in the inset in Fig. \ref{Figure4} is indeed an indirect test of the vertical homogeneity of the films, as discussed in details in Ref. \cite{SupplementaryMaterial}.

We now come to a discussion of our results. First, note that the thickness dependence of $SR$ clearly demonstrates that the conductance relaxation exists in the "bulk"" of the film and is a property of the granular $Al$ itself (and not of its interface with the gate or the surface oxide layer). The "bulk" relaxation after a cool down at $4.2K$ is well described by a $\ln t$ dependence without any sign of saturation over weeks of measurement. Such a $\ln t$ dependence was also observed for the time increase of the conductance dip amplitude after a cool down \cite{OvadyahuPRB02,GrenetEPJB07} (more complex laws are observed when $V_{geq}$ is fixed a time $t_e$ after the cool down, see \cite{GrenetAging10}). It is a natural relaxation law for a system having a $1/\tau$ distribution of relaxation times, as expected for an electron glass \cite{AmirRecent}.

In our granular $Al$ thin films, the electronic transport at finite $T$ is believed to result from electron tunneling between metallic $Al$ grains. A lower bound for $L_{sc}$ is then the typical grain size, which was found in recent TEM measurements to be $2-4nm$. A second important microscopic length scale is the percolation radius $L_0$ of the critical resistance network. Since in our films $R_C \lesssim 100G\Omega$, the diffusion of charge carriers through the critical resistance network is expected to be faster than the time scale of the measurements (the intergrain capacitance and the capacitance of the granular $Al$ channel to the gate are respectively $ \simeq 10^{-19}F$ and $\simeq 10^{-11}F$). Thus, $L_0$ should be an upper bound for $L_{sc}$ as indeed observed: $L_0$ estimates from conductance fluctuations measurements are between $20nm$ and $40nm$ \cite{DelahayeEPJB08}, a few times our $L_{sc}$ value.

We note that no baseline relaxation was reported on $20nm$ thick indium oxide films which may indicate that $L_{sc}(4K)>20nm$ in this system. Interestingly enough, $L_0$ was found to be $\simeq 300nm$ in "crystalline" films \cite{OrylanchikPRB07}, i.e. $\simeq 10$ times our estimate for granular $Al$ films, reflecting a less dense percolating network in indium oxide consistent with a larger screening length.

Our observations address the important question of screening in a disordered insulator, a subject which has been little explored both theoretically and experimentally. Electron glass models are generally developed in the limit of strongly localized electrons \cite{DaviesPRL82,ElectronGlass80,RecentEGTheory} (Ref. \cite{VignalePRB87,Dobrosavljevic} are exceptions). But in real systems, there is a mobility of the charge carriers at finite $T$ which will give rise to a metallic-like screening. Numerical studies on disordered insulators have indeed found a transition between a metallic screening at high $T$ where most of the electrons are diffusive and a slow dielectric response at low $T$ where most of the electrons remain located in finite size clusters \cite{XuePRL88,DiazPRB99,KoltonPRB05}. Such a transition was observed around $1K$ in capacitive measurements on a doped semiconductor \cite{DonMonroePRL87}. According to the usual formula, $L_{sc}^2 = \epsilon \epsilon_0 /e^2 dN/d\mu(E_F)$, where $dN/d\mu$ is the thermodynamic DOS. In a glassy phase and for a finite time $t$, the system cannot explore all the configurations and the thermodynamic DOS has to be replaced by a pseudo-equilibrium DOS $dN/d\mu(t)$ for which only relaxations faster than $t$ are allowed. The first theoretical attempts to do so \cite{DaviesPRL82,MullerPRB07} have found that the short time ($\simeq$ Maxwell time) screening length diverges as $T$ goes to 0 and decreases as a function of time \cite{MullerPRB07}.

In summary, we have described in insulating granular $Al$ films the existence of a $V_g$ insensitive conductance relaxation. Its thickness dependence demonstrates the existence of a metallic screening length of about $10nm$ at $4K$. Our results provide a new way to study the screening length and its relaxation in disordered insulators where an anomalous electrical field effect has been found. They also point to the need for more theoretical studies: how are the classical electron glass models predictions affected by the existence of a metallic screening length ?

This research has been partly supported by the French-National Research Agency ANR (contract $N^o 05-JC05-44044$).




\begin{thebibliography}{0}
\expandafter\ifx\csname natexlab\endcsname\relax\def\natexlab#1{#1}\fi
\expandafter\ifx\csname bibnamefont\endcsname\relax
  \def\bibnamefont#1{#1}\fi
\expandafter\ifx\csname bibfnamefont\endcsname\relax
  \def\bibfnamefont#1{#1}\fi
\expandafter\ifx\csname citenamefont\endcsname\relax
  \def\citenamefont#1{#1}\fi
\expandafter\ifx\csname url\endcsname\relax
  \def\url#1{\texttt{#1}}\fi
\expandafter\ifx\csname urlprefix\endcsname\relax\def\urlprefix{URL }\fi
\providecommand{\bibinfo}[2]{#2}
\providecommand{\eprint}[2][]{\url{#2}}

\end{thebibliography}


\begin{thebibliography}{99}

\bibitem{FirstExpElectronGlass} M. Ben-Chorin \emph{et al.}, Phys. Rev. B \textbf{44}, 3420 (1991); G.
Martinez-Arizala \emph{et al.}, Phys. Rev. Lett. \textbf{78}, 1130 (1997); T. Grenet, Eur. Phys. J. B \textbf{32}, 275 (2003).

\bibitem{OvadyahuPRB02} A. Vaknin \emph{et al.}, Phys. Rev. B \textbf{65}, 134208 (2002).

\bibitem{GrenetEPJB07} T. Grenet \emph{et al.}, Eur. Phys. J. B  \textbf{56}, 183 (2007).

\bibitem{GrenetAging10} T. Grenet \emph{et al.}, to be published in Eur. Phys. J. B.

\bibitem{OvadyahuPRL98} A. Vaknin \emph{et al.}, Phys. Rev. Lett. \textbf{81}, 669 (1998).

\bibitem{OvadyahuElectronGlass} Z. Ovadyahu, Phys. Rev. Lett. \textbf{99}, 226603 (2007); Z. Ovadyahu, Phys. Rev. B \textbf{78}, 195120 (2008); Z. Ovadyahu, Phys. Rev. Lett. \textbf{102}, 206601 (2009).

\bibitem{DaviesPRL82} J. H. Davies \emph{et al.}, Phys. Rev. Lett. \textbf{49}, 758 (1982).

\bibitem{ElectronGlass80} M. Gr\"unewald \emph{et al.}, J. Phys. C \textbf{15}, L1153 (1982); M. Pollak and M. Ortu\~no, Sol. Energy Mater. \textbf{8}, 81 (1982).

\bibitem{RecentEGTheory} For recent Ref., see
C.C. Yu, Phys. Rev. Lett. \textbf{82}, 4074 (1999);
D.N. Tsigankov \emph{et al.}, Phys. Rev. B \textbf{68} 184205 (2003).
M. M\"uller and L. Ioffe, Phys. Rev. Lett. \textbf{93}, 256403 (2004);
V. Malik and D. Kumar, Phys. Rev. B \textbf{69}, 153103 (2004);
D.R. Grempel, Europhys. Lett. \textbf{66}, 854 (2004);
E. Lebanon and M. M\"uller, Phys. Rev. B \textbf{72}, 174202 (2005);
A.M. Samoza \emph{et al.}, Phys. Rev. Lett. \textbf{101}, 056601 (2008).



\bibitem{SupplementaryMaterial} See EPAPS Document $N^o$ for a discussion of the vertical electrical homogeneity of the granular $Al$ films.

\bibitem{ESBook84} B.I. Shklovskii and A.L. Efros, E\emph{lectronic Properties of Doped
Semiconductors} (Springer-Verlag, New York 1984).

\bibitem{DelahayeEPJB08} J. Delahaye \emph{et al.}, Eur. Phys. J. B \textbf{65}, 5 (2008). The existence of a screening length was not taken into account in this Ref. and the $L_0$ values reported for $20nm$ films must be divided by a factor $T_h/L_{sc}\simeq \sqrt2$.

\bibitem{AmirRecent} A. Amir \emph{et al.}, Phys. Rev. B \textbf{77}, 165207 (2008); A. Amir \emph{et al.}, Phys. Rev. Lett. \textbf{103}, 126403 (2009).



\bibitem{OrylanchikPRB07} V. Orlyanchik and Z. Ovadyahu Phys. Rev. B \textbf{75}, 174205 (2007).


\bibitem{VignalePRB87} G. Vignale, Phys. Rev. B \textbf{36}, 8192 (1987).

\bibitem{Dobrosavljevic}  A.A. Pastor and V. Dobrosavljevi\'c, Phys. Rev. Lett. \textbf{83}, 4642 (1999); V. Dobrosavljevi\'c \emph{et al.}, Phys. Rev. Lett. \textbf{90}, 016402 (2003).

\bibitem{XuePRL88} W. Xue and P.A. Lee, Phys. Rev. B \textbf{38}, 9093 (1988).

\bibitem{KoltonPRB05} A.B. Kolton \emph{et al.}, Phys. Rev. B \textbf{71}, 024206 (2005).

\bibitem{DiazPRB99} A. D\'iaz-S\'anchez \emph{et al.}, Phys. Rev. B \textbf{59}, 910 (1999).

\bibitem{DonMonroePRL87} Don Monroe \emph{et al.}, Phys. Rev. Lett. \textbf{59}, 1148 (1987).

\bibitem{MullerPRB07} M. M\"uller and S. Pankov, Phys. Rev. B \textbf{75}, 144201 (2007).




\end{thebibliography}
\end{document}